\documentclass[prb,showpacs,twocolumn]{revtex4}
\usepackage{graphicx}
\usepackage{dcolumn}
\usepackage{bm}
\usepackage{amssymb}
\begin{document}
\input psfig.tex
\def\be{\begin{equation}}
\def\ee{\end{equation}}
\def\ba{\begin{eqnarray}}
\def\ea{\end{eqnarray}}
\newcommand{\s}{\sigma}
\title{Local field factors in a polarized two-dimensional electron gas}

\author{Juana Moreno and D. C. Marinescu}
\affiliation{Department of Physics, Clemson University, Clemson, SC 29634}
\date{\today}
\begin{abstract}

We derive approximate expressions for the static local field factors 
of a spin polarized two-dimensional electron gas
which smoothly interpolate
between their small- and  large-wavevector asymptotic limits.
For the unpolarized electron gas, the proposed
analytical expressions  reproduce recent diffusion
Monte Carlo data. We find that the degree of spin polarization produces
important modifications to the local factors of the minority spins,
while the local field functions of the majority spins are less affected.

\end{abstract}

\maketitle

\section{\bf Introduction}

In most of 
the many body theories of electronic systems, the spin
polarization in a magnetic field has been 
considered as a small parameter. This
picture is certainly true when the Zeeman splitting
is much smaller than  other relevant energies in the
problem. 
However, recently, attention has been focused on materials
where the Zeeman splitting dominates the energy spectrum, 
such as the diluted magnetic semiconductors. 
The large Zeeman splitting is due to the strong exchange 
interactions between the itinerant carriers and the magnetic
ions that generate a value of the effective gyromagnetic factor,
$\gamma^*$, up to hundreds of times its band value. 
As a consequence, even in weak magnetic fields these systems can be
fully polarized.\cite{crooker95} Due to their easy polarizability
these materials are promising candidates to 
accomplish spin dependent conduction in solid state devices;\cite{review}
hence the interest in developing an appropriate microscopic model
for them.

A realistic picture of the spin polarized electron gas hinges on
finding an appropriate description of the many-body interaction,
which has to  incorporate
the explicit spin dependence of the short range Coulomb repulsion. 
In this paper, we examine these short range effects 
in a two dimensional polarized  electron gas.
The self-consistent treatment of the exchange and correlation effects
has proven to be very important in understanding the physics of normal
metals, \cite{mahan} but to our knowledge it has not been fully
analyzed in spin-polarized systems. 
In addition, the relevance of the exchange and correlation 
effects increases as the dimensionality of the electron gas is lowered.

We  model
the exchange (x) and correlation (c) hole around each electron
by using spin dependent local field
correction functions, $G^{x,c}_{\sigma}({\bf q},\omega)$.\cite{yi96}
They describe the
difference between  the true particle density and its mean field (RPA)
counterpart. The spin dependence of these factors is a consequence of
the fact that the microscopic interaction 
is different for up and down spin electrons.
The microscopic origin of the local field corrections was
elucidated by Kukkonen and Overhauser. \cite{kuk79} 
In a direct generalization of this approach, the effective potential
$\widetilde{\cal V}_{\sigma}$ experienced by an electron
with spin $\sigma$ in the presence of 
external electric
[$\phi(\bf q, \omega)$] and magnetic fields [$\vec B(\bf q, \omega)$]
can be written as:\cite{yi96,zhu86}
\ba
\widetilde{\cal V}_{\sigma}({\bf q}, \omega) =  -e \phi({\bf q}, \omega) +
\gamma^{*} \vec{\sigma} \cdot \vec{B}({\bf q}, \omega) + \nonumber 
\ea
\be
v({\bf q}) \Big\{ [1 - \widetilde{G}^{+}_{\sigma}({\bf q},\omega)] 
\Delta n({\bf q},\omega)
- \vec{\sigma} \cdot \Delta \vec{m}({\bf q},\omega)
\widetilde{G}^{-}_{\sigma}({\bf q},\omega) \Big\}
\label{eq:eff.int}
\ee
where
$\Delta n= \Delta n_{\uparrow} +\Delta n_{\downarrow}$ 
and $\Delta \vec m$ are, respectively, the density and 
the magnetization fluctuation  induced by the external fields,
$\vec{\sigma}$ are the Pauli
spin matrices 
and $v({\bf q})$ is the bare Coulomb interaction, $2 \pi e^2/{\bf q}$.
The local field factor $\widetilde{G}^{+}_{\sigma}({\bf q},\omega)$
is the sum of parallel- and antiparallel-spin effects,
while 
$\widetilde{G}^{-}_{\sigma}({\bf q},\omega)$ is the difference of 
parallel- and antiparallel-spin effects: 
$\widetilde{G}^{\pm}_{\sigma}({\bf q},\omega)=
\widetilde{G}^x_{\sigma}+\widetilde{G}^c_{\sigma \sigma} \pm 
\widetilde{G}^c_{\sigma \bar{\sigma}}$. 
\cite{unpol-pol}

In a linear approximation, the density fluctuations are
proportional to the effective potentials:
 $\Delta n_{\uparrow}= \Pi_{ \uparrow
\uparrow} \widetilde{\cal V}_{\uparrow}$ ($\Delta
n_{\downarrow}= \Pi_{ \downarrow \downarrow} \widetilde {\cal
V}_{\downarrow}$),\cite{yi96, mar97}
where the proportionality coefficients are 
the polarization functions of the fully interacting electron
system.
Since the interacting polarization function is 
generally unknown,  an additional
local field factor ($G^{n}_{\sigma}$) is needed to relate
the interacting polarization function, 
$\Pi_{\sigma\sigma}$, with the
non-interacting one, $\Pi^0_{\sigma\sigma}$,  as  in\cite{Richar94-97}
\be
\Pi_{\sigma\sigma}({\bf q}, \omega)=
\frac{\Pi^0_{\sigma\sigma}({\bf q}, \omega)}{1+ 2 v({\bf q}) 
G^{n}_{\sigma}({\bf q}, \omega) \Pi^0_{\sigma\sigma}({\bf q}, \omega)}
\label{eq:intLindhard}
\ee

This parameterization of the modified polarization function
leads to a renormalized expression  of the local fields
that determine the response functions, such that 
the complete local field functions are given as:
$G^{\pm}_{\sigma}= \widetilde{G}^{\pm}_{\sigma}+G^{n}_{\sigma}$.
\cite{Richar94-97}

Quantitative calculations of many physical properties 
require the precise knowledge of the local
field correction functions. The determination of 
the frequency and wave vector dependence of the
local field corrections is a very difficult problem which
remains unsolved even in the case of the unpolarized
electron system. Fortunately, the asymptotic values
of the local field factors 
can be obtained exactly in some limiting cases.\cite{mar97,mar02,pol01}
Numerical estimates of the response functions of the two and three
dimensional unpolarized electron gas have shown that local field
factors smoothly interpolate between the asymptotic small and
large wave-vector behavior.\cite{DMC} 
This feature is expected 
to exist also in the case of a spin polarized system, and,
consequently, we use the asymptotic limits of
the static local factors for large
and small wavevector
as a starting point in
deriving their approximate expressions across the whole spectrum
of momentum.

The fundamental parameters of the problem are 
the coupling strength,
$\displaystyle r_s=a^{*}_B/\sqrt{\pi (n_{\uparrow} +n_{\downarrow})}=
a^{*}_B/\sqrt{\pi n}$,\cite{numbers}
and the spin polarization,
$\displaystyle \zeta= (n_{\uparrow}-n_{\downarrow})/n$.
Since the many body 
interaction is independent of the source of the polarization 
we expect our results to maintain their validity also in 
the case of an itinerant ferromagnet with a self-induced magnetic field, 
or when the polarization
is achieved by other means, such as shining circularly polarized 
light on the sample.

In section II, we study the large and the small wavevector
limits of the  local field functions and their dependence 
with the electronic density and  polarization.
In section III, we give a simple parameterization of the local 
field factors which satisfy the asymptotic limits and 
reproduces the most recent numerical results
for the unpolarized electron gas. \cite{DMC}
Section IV presents our conclusions.

\section{Limiting behavior of the local field factors}

Our first approximation is
to neglect the frequency dependence of the local field
corrections. Although the local field functions represent a
dynamical effect,\cite{Richar94-97,Atwal02} 
they vary slowly on the scale of the Fermi
frequency,\cite{hol79,pede97} and it is acceptable to neglect
their frequency dependence if we are mainly interested in response
functions. 

\subsection{Small wavevector}

At zero frequency
and small wavevector, sum rules are used to connect the static
limits of various response functions to certain thermodynamic
coefficients, which can be  expressed as  
derivatives of the ground state energy of the electron gas.
\cite{mar02} Subsequently,
the renormalized local field functions, which are directly connected
with these response functions,
are written down as derivatives of the exchange and correlation
energy  of the interacting electron gas ($E^{xc}$):
\be
G^{+}_{\sigma}({\bf q}\rightarrow 0)
= \frac {\tilde{q} r^2_s}{8 \sqrt{2}}
\left(\frac{\partial\epsilon^{xc}}{\partial r_s}
- r_s \frac{\partial^2 \epsilon^{xc}}{\partial r^2_s}
+ 2~ sign (\sigma) \frac{\partial^2 \epsilon^{xc}} 
{\partial r_s \partial \zeta} \right), 
\ee
\be
G^{-}_{\sigma}({\bf q}\rightarrow 0)= 
\frac {\tilde{q} r_s}{2 \sqrt{2}}
\left(
-  \frac{\partial^2 \epsilon^{xc}}{\partial \zeta^2}
+ sign (\sigma) \frac{r_s}{2}  \frac{\partial^2 \epsilon^{xc}} 
{\partial r_s \partial \zeta} \right), 
\ee
where $\tilde{q}={\bf q}/k_F$ is the  normalized momentum,
$\epsilon^{xc}=E^{xc}/N$ is the exchange and correlation
energy per particle measured in Rydbergs and the coupling strength,
$r_s$, is measured in units of the effective Bohr radius of the
system. \cite{numbers} 
Using the explicit expression of the 
exchange energy the local field functions become:
\ba
G^{+}_{\uparrow}({\bf q}\rightarrow 0)&&= \frac {\tilde{q}}{2 \pi}
\left( (2+\zeta) \sqrt{1+\zeta} -\zeta \sqrt{1-\zeta} \right)\nonumber \\
&&+\frac {\tilde{q} r^2_s}{8 \sqrt{2}}
\left(\frac{\partial\epsilon^{c}}{\partial r_s}
- r_s \frac{\partial^2 \epsilon^{c}}{\partial r^2_s}
+ 2  \frac{\partial^2 \epsilon^{c}} 
{\partial r_s \partial \zeta} \right), \\
G^{-}_{\uparrow}({\bf q}\rightarrow 0)&&= \frac {\tilde{q}}{2 \pi}
\left( \frac{2+\zeta}{ \sqrt{1+\zeta}}+ \frac{\zeta} 
{\sqrt{1-\zeta}} \right) \nonumber \\
&&+\frac {\tilde{q} r_s}{2 \sqrt{2}}\left(
-  \frac{\partial^2 \epsilon^{c}}{\partial \zeta^2}
+  \frac{r_s}{2}  \frac{\partial^2 \epsilon^{c}} 
{\partial r_s \partial \zeta} \right). 
\ea
Similar expressions apply to $G^{\pm}_{\downarrow}$ due 
to the fact that 
$G^{\pm}_{\downarrow}(\zeta)=
G^{\pm}_{\uparrow}(-\zeta)$.

\begin{figure}
\begin{minipage}{\linewidth}
\includegraphics[width=\textwidth]{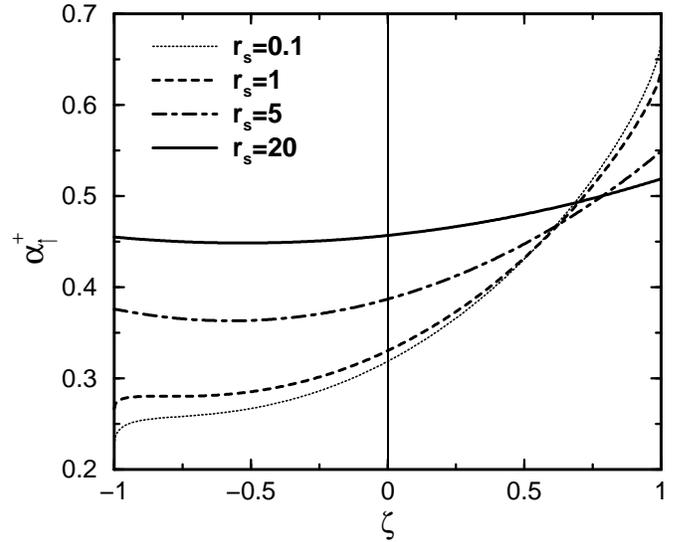}
\caption {Initial slope $\alpha^{+}_{\uparrow}(r_s,\zeta)$ 
of the local field factor
as a function of the polarization,
$\zeta$, for $r_s=0.1$ (dotted line), $r_s=1.$ (dashed line),
$r_s=5.$ (dot-dashed line) and $r_s=20.$ (solid line).}
\label{fig:alphapluspol}
\end{minipage}
\end{figure}

\begin{figure}
\begin{minipage}{\linewidth}
\includegraphics[width=\textwidth]{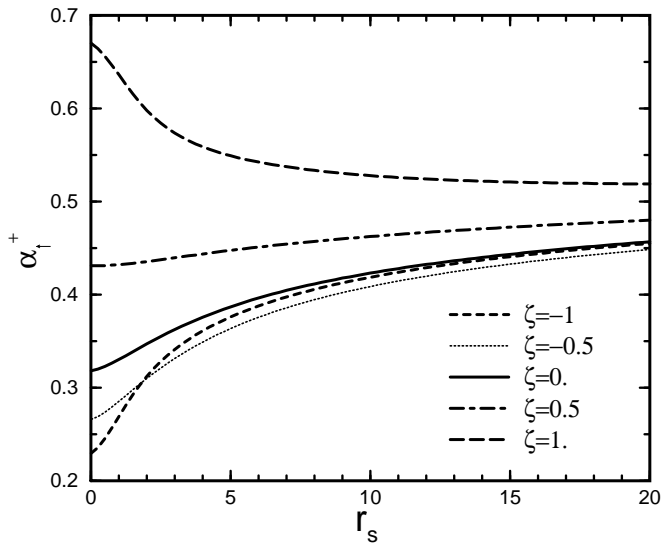}
\caption {Initial slope $\alpha^{+}_{\uparrow}(r_s,\zeta)$ of the local 
field correction as a function of the coupling strength
$r_s$ for different values of the spin polarization:
$\zeta=-1$ (dashed line), $\zeta=-0.5$ (dotted line),
$\zeta=0.$ (solid line), $\zeta=0.5$ (dot-dashed line)
and $\zeta=1$ (long-dashed line).}
\label{fig:alphaplusrs}
\end{minipage}
\end{figure}

The  inclusion of  the correlation energy 
in the small $\bf q$ limit
of the local field factors is crucial 
to correctly evaluate the response functions.
If only the exchange contribution to
the local field factors is included, the 
magnetic susceptibility of the unpolarized gas
develops a pole for $r_s \ge \pi/\sqrt{2}\sim
2.221$, \cite{dielectric} in disagreement with the results
of extensive numerical calculations that 
show a stable paramagnetic phase up to $r_s
\lesssim 26$.\cite{Tan89,Moroni02} 
The addition of
the correlation energy in 
the the small ${\bf q}$ limit of the local field factors
prevents the occurrence of an instability in the
unpolarized electron gas.

To precisely evaluate the contribution from  the correlation energy 
to $G^{\pm}_{\sigma}({\bf q}\rightarrow 0)$
we use the latest numerical computation of the ground-state
energy of the two-dimensional electron gas. Following
Attaccalite et {\it al.}, \cite{Moroni02}
the correlation energy per particle 
is given by: 
\ba
\epsilon_c (r_s, \zeta)= (e^{-\beta r_s} -1 ) 
\epsilon^{(6)}_x (r_s, \zeta)  \nonumber \\
+ \alpha_0(r_s) + \alpha_1(r_s) \zeta^2 +  \alpha_2(r_s) \zeta^4,
\label{eq:correnergy}
\ea
where $\epsilon^{(6)}_x$ is the Taylor expansion of the exchange
energy beyond fourth order in $\zeta$, $\beta$ is a parameter and
the functions $\alpha_i(r_s)$ are a generalization of previous
expressions. \cite{Perdew92}
By taking the appropriate derivatives of  
Eq.~(\ref{eq:correnergy}), we
derive analytical expressions for  the initial slope of the static
local field correction functions:
$\alpha^{\pm}_{\sigma}=G^{\pm}_{\sigma}(\tilde{q}\rightarrow
0)/\tilde{q}$.

The behavior of $\alpha^{+}_{\uparrow}(r_s,\zeta)$
and $\alpha^{-}_{\uparrow}(r_s,\zeta)$ is quite different.\cite{similar} 
Fig.~\ref{fig:alphapluspol}
shows $\alpha^{+}_{\uparrow}(r_s,\zeta)$ as a
function of the polarization for several values of $r_s$. Note
that $\alpha^{+}_{\uparrow}$ is always positive, as the
local effects always decrease the uniform electron 
density at large inter-electronic distances. Also note that
$\alpha^{+}_{\uparrow}$ for small values of $r_s$ 
is a  monotonically increasing function 
of $\zeta$. This behavior can be understood analyzing the 
contribution from the exchange interaction given by  
$\alpha^{+}_x=[ (2+\zeta) 
\sqrt{1+\zeta} -\zeta \sqrt{1-\zeta} ]/2 \pi$. 
Since the exchange interaction
takes place only between electrons with parallel spin,
increasing values of $\zeta$ induce larger  effects 
and further reduction of the uniform electron 
density at small ${\bf q}$. Consequently, the value of
$\alpha^{+}_{\uparrow}$ increases. The exchange contribution $\alpha^{+}_x$, 
which does not depend on $r_s$, dominates at large electronic 
densities. However, correlation effects become more important 
with increasing $r_s$ and they partly cancel the strong $\zeta$ dependence 
of $\alpha^{+}_x$, as it can be seen on Fig.~\ref{fig:alphapluspol}.

\begin{figure}
\begin{minipage}{\linewidth}
\includegraphics[width=\textwidth]{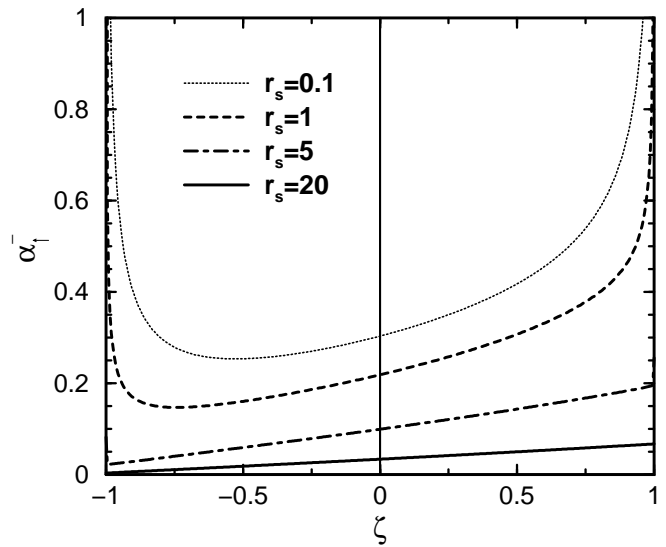}
\caption {Initial slope $\alpha^{-}_{\uparrow}(r_s,\zeta)$ 
of the local field factor
as a function of the polarization
for different values of $r_s$ as indicated in the legend.} 
\label{fig:alphaminuspol}
\end{minipage}
\end{figure}

\begin{figure}
\begin{minipage}{\linewidth}
\includegraphics[width=\textwidth]{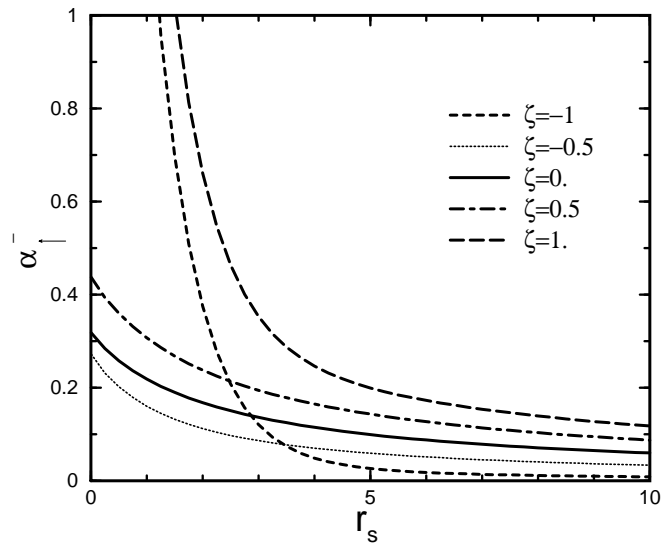}
\caption {Initial slope $\alpha^{-}_{\uparrow}(r_s,\zeta)$
as a function of $r_s$ for different values of the spin polarization.} 
\label{fig:alphaminusrs}
\end{minipage}
\end{figure}

Fig.~\ref{fig:alphaplusrs} displays $\alpha^{+}_{\uparrow}(r_s,\zeta)$ 
as a function of $r_s$ for different values of the spin polarization. 
It can be noticed that $\alpha^{+}_{\uparrow}$ is a monotonic increasing
function of $r_s$ for values of  $\zeta \lesssim 0.5$.  
This feature is also displayed on Fig.~\ref{fig:alphapluspol}.
In particular, as previously noticed,\cite{DMC}  
in the unpolarized electron gas
$\alpha^{+}_{\uparrow}(r_s,\zeta=0)$ increases with $r_s$.

Fig.~\ref{fig:alphaminuspol} displays the initial slope 
$\alpha^{-}_{\uparrow}(r_s,\zeta)$ as a function of $\zeta$
for different values of $r_s$. As in the case of
$\alpha^{+}_{\uparrow}$,  the main contribution
for small values of $r_s$ came from the exchange energy:
$\displaystyle  \alpha^{-}_x=\frac{1}{2 \pi}
\left( \frac{2+\zeta}{ \sqrt{1+\zeta}}+ \frac{\zeta} 
{\sqrt{1-\zeta}}\right)$. Therefore, for large electronic densities 
$\alpha^{-}_{\uparrow}$ diverges for a fully polarized system 
($\zeta= \pm 1$), behavior explained by the fact that 
in a fully polarized gas the magnetic susceptibility becomes zero.
As $r_s$ increases, correlation effects quench 
by a factor of $e^{-\beta r_s}$ 
the diverging contribution from the exchange energy.\cite{approximation}
Thus, correlation
effects become dominant at small densities and the strong dependence
of $\alpha^{-}_x$ with $\zeta$ is completely washed out. Instead, at
large values of $r_s$, $\alpha^{-}_{\uparrow}$ becomes a linear function 
of $\zeta$.

Fig.~\ref{fig:alphaminusrs} displays $\alpha^{-}_{\uparrow}(r_s,\zeta)$ 
as a function of $r_s$ for different values of the spin polarization. 
The divergence at $r_s=0$ and $\zeta = \pm 1$ is clearly displayed.
Note that $\alpha^{-}_{\uparrow}$ is a monotonic decreasing
function of $r_s$ for any value of  $\zeta$.

\subsection{Large wavevector}

In the limit of large wavevector it is easier to derive 
independently expressions for $\widetilde{G}^{\pm}_{\sigma}$ and 
for $G^{n}_{\sigma}$. For large frequency or large
wavevector, an iterative method generates the exact 
asymptotic expressions for  $\widetilde{G}^{\pm}_{\sigma}$. \cite{niklasson}
For a two-dimensional electron gas, \cite{pol01}
\ba
\widetilde{G}^{+}_{\uparrow}({\bf q}\rightarrow \infty)= 
\beta^{+}_{\uparrow}(r_s, \zeta)=
1 -\frac{1}{2}g_{\uparrow \downarrow}(0)\\
\widetilde{G}^{-}_{\uparrow}({\bf q}\rightarrow \infty)= 
\beta^{-}_{\uparrow}(r_s, \zeta)=
\frac{1}{2}g_{\uparrow \downarrow}(0)
\ea
where $g_{\uparrow \downarrow}(0)$ is the spin resolved pair
distribution function at the origin. It has been shown that
$g_{\uparrow \downarrow}(0)$ is largely unaffected by the degree
of spin polarization. \cite{Polini01,Moreno02} 
In our calculation we use the simple  expression 
$\displaystyle g_{\uparrow \downarrow}(0)=
\frac{1}{[1+0.6032~ r_s +0.07263~ r_s^2]^2}$,
where only the parameter $r_s$ appears. \cite{Moreno02}
With this choice, $\beta^{+}_{\uparrow}(r_s, \zeta)=
\beta^{+}_{\downarrow}(r_s, \zeta)=\beta^{+}(r_s)$ and
$\beta^{-}_{\uparrow}(r_s, \zeta)=
\beta^{-}_{\downarrow}(r_s, \zeta)=\beta^{-}(r_s)$.

In order to calculate the large momentum behavior of
$G^{n}_{\sigma}$ we need to approximate the modified 
polarization function,\cite{Holas87,Richar94-97}
\be
\Pi_{\sigma \sigma'}({\bf q},\omega)= \frac{1}{A}
\sum_{\bf k} \frac{\bar{n}_{\bf k, \sigma} - \bar{n}_{\bf k +\bf q, \sigma'}}
{\omega + \xi_{\bf k \sigma} - \xi_{\bf k+ \bf q \sigma'}}
\label{eq:polarizationbuble}
\ee
where $\xi_{\bf k \sigma}=
\epsilon_{\bf k} + \mbox{sign}(\sigma) \gamma^* B=
k^2/2 m + \mbox{sign}(\sigma) \gamma^* B$ is the 
quasiparticle energy in the static magnetic field $B$,
$\bar{n}_{\bf k, \sigma}$ is
the exact occupation numbers in the interacting electron gas,
and A is the area of the system. Since
at large ${\bf q}$ the 
particle number renormalization is the dominant effect, 
Eq.~(\ref{eq:polarizationbuble}) neglects 
the renormalization of the quasiparticle effective mass.

By making an asymptotic expansion of the interacting polarization 
function,\cite{Holas87}, the local factor $G^{n}_{\sigma}$ 
can be written down as,\cite{Davoudi01,Giuliani-book} 
\ba
G^{n}_{\sigma}({\bf q}\rightarrow \infty)=
\frac{r_s}{\sqrt{2}}\frac{\Delta t_{\sigma}}{(1+\zeta)^2}
\tilde{q} = \gamma_{\sigma}(r_s, \zeta) \tilde{q}
\label{eq:Gn}
\ea
where $\tilde{q}={\bf q}/k_F$ is the  normalized momentum,
$\widetilde{k}_{F\sigma}=k_{F\sigma}/k_F$
the normalized Fermi momentum of the spin $\sigma$ electronic population
and $\Delta t_{\sigma}=t_{\sigma} -t^0_{\sigma}$ is the difference between 
the kinetic energy of the electrons with spin $\sigma$ 
in the interacting system,
$\displaystyle t_{\sigma}= \frac{1}{N}\sum_k \bar{n}_{k \sigma} 
\frac{k^2}{2 m}$,
 and in the non-interacting gas, 
$\displaystyle t^0_{\sigma}= \frac{1}{N}\sum_k n^0_{k \sigma}
\frac{k^2}{2 m}$, over the total number of electrons. This equation is valid when 
$\Delta t_{\sigma}$ is measured in Rydbergs and $r_s$ in 
units of the effective Bohr radius. \cite{numbers}

The difference in kinetic energies can be related with 
the exchange and correlation energies and their derivatives 
using the virial\cite{March58} and the magnetic virial theorems: \cite{pol01}
\be
\Delta t_{\sigma}= -\frac{r_s}{2} \frac{\partial(\epsilon^x +\epsilon^c)}
{\partial r_s} - \epsilon^x_{\sigma} -\epsilon^c_{\sigma}
\ee  
where $\epsilon^x= E^x/N$ and $\epsilon^c= E^c/N$
are, respectively, the average exchange and correlation 
energy per particle; $\epsilon^x_{\sigma}=E^x_{\sigma}/N$
and $\epsilon^c_{\sigma}=E^c_{\sigma}/N$ are the average
exchange and correlation energy of the electrons with spin $\sigma$.

The average exchange energy for any spin population is well known:
$\displaystyle \epsilon^x_{\sigma}=-\frac{4 \sqrt{2}}{3 \pi r_s}
(1+\mbox{sign}(\sigma) \zeta)^{3/2}$ (Ry). The spin dependent
correlation energies are difficult to evaluate or to extract 
from numerical calculations. Therefore, we have to rely on
same approximate scheme  to extract the spin 
dependent correlation energy from the available computations
of the full correlation energy.
The total correlation energy per particle is 
$\displaystyle \frac{E^c}{N}=\epsilon^c_{\uparrow}+
\epsilon^c_{\downarrow}
=\widetilde{\epsilon}^c_{\uparrow}\frac{1+\zeta}{2}
+\widetilde{\epsilon}^c_{\downarrow}\frac{1-\zeta}{2}$. 
Perdew and Wang \cite{Perdew92} 
suggested the following parameterization for the correlation energy
of a polarized electron gas:
\be
\epsilon^c(r_s,\zeta)=\epsilon^c(r_s,0)+ h(r_s,\zeta) f(\zeta)
\ee
where $ h(r_s,\zeta)$ is an even function of the polarization and
$\displaystyle f(\zeta)= \frac{(1+\zeta)^{3/2}+(1-\zeta)^{3/2}-2}
{2(\sqrt{2}-1)}$ for a two-dimensional system.\cite{DMC}
This function can be decomposed
as $\displaystyle f(\zeta)=[f_{\uparrow}(\zeta)(1+\zeta)+
f_{\downarrow}(\zeta)(1-\zeta)]/2$, where 
$f_{\sigma}=[\sqrt{1+sign(\sigma) \zeta} -1]/[\sqrt{2}-1]$.
An estimate of the spin dependent correlation energies can be
obtained as: 
\ba
\epsilon^c_{\sigma}(r_s,\zeta)=\frac{1+sign(\sigma) \zeta}{2}
~\widetilde{\epsilon}^c_{\sigma}(r_s,\zeta),\hspace{0.2in} {\rm where}
\nonumber \\
 \widetilde{\epsilon}^c_{\sigma}(r_s,\zeta)=\epsilon^c(r_s,0)+ 
\frac{f_{\sigma}(\zeta)}{f(\zeta)}
[\epsilon^c(r_s,\zeta)-\epsilon^c(r_s,0)].
\label{eq:spincorrelation}
\ea

Using the parameterization
of the correlation energy proposed in Ref.~\onlinecite{Moroni02}, 
Eq.~(\ref{eq:correnergy}), we obtain
reasonable values for $\epsilon^c_{\sigma}$. Also, its dependence
with the polarization is the expected one. At a fixed value of
$r_s$, $\epsilon^c_{\uparrow}$
($\epsilon^c_{\downarrow}$)
is a monotonic increasing (decreasing) function of $\zeta$.

\begin{figure}
\begin{minipage}{\linewidth}
\includegraphics[width=\textwidth]{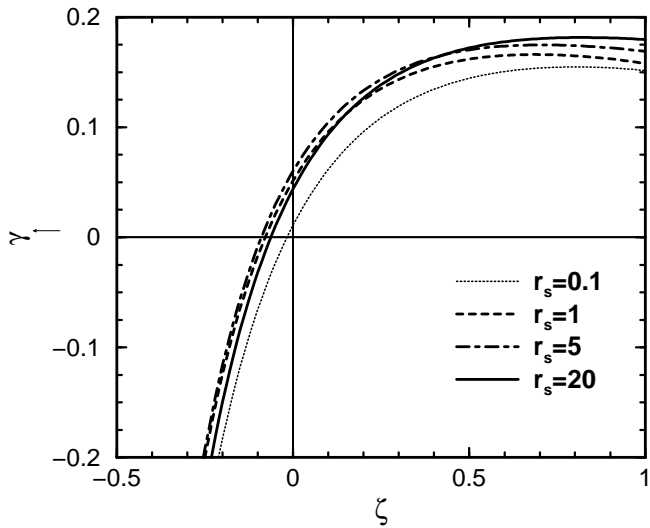}
\caption {The coefficient $\gamma_{\uparrow}(r_s,\zeta)$
as a function of 
$\zeta$ for different values of $r_s$ as indicated in the
legend.} 
\label{fig:gammapol}
\end{minipage}
\end{figure}

\begin{figure}
\begin{minipage}{\linewidth}
\includegraphics[width=\textwidth]{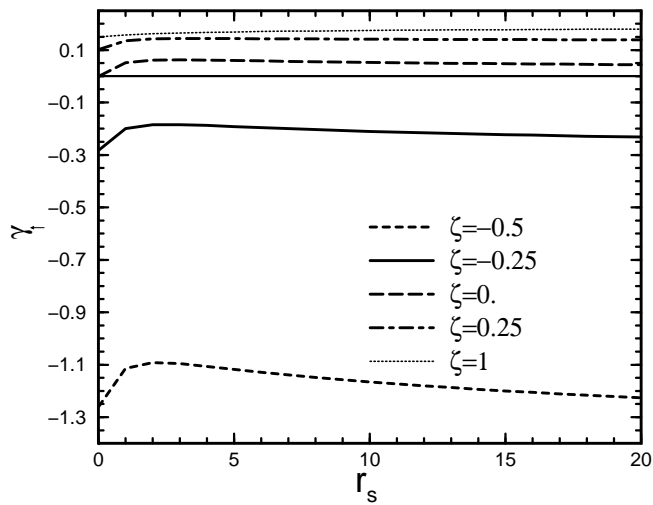}
\caption {The coefficient $\gamma_{\uparrow}(r_s,\zeta)$
as a function of the coupling strength $r_s$
for different values of the polarization:
$\zeta=-0.5$ (dashed line), $\zeta=-0.25$ (solid line),
$\zeta=0.$ (long-dashed line), $\zeta=0.25$ (dot-dashed line)
and $\zeta=1$ (dotted line).}
\label{fig:gammars}
\end{minipage}
\end{figure}

Fig.~\ref{fig:gammapol} shows the dependence of 
$ \gamma_{\uparrow}(r_s,\zeta)$, Eq.~(\ref{eq:Gn}),
with the polarization for several values of $r_s$.
While the initial slopes ($\alpha^{\pm}_{\sigma}$)
and the constant terms in the large ${\bf q}$ limit 
($\beta^{\pm}$) of  the local  factors
are always positive, 
the coefficient $\gamma_{\sigma}$
can have any sign. Since  $\gamma_{\sigma} \propto \Delta t_{\sigma}$,
a change in sign of $\gamma_{\sigma}$ implies a change of  sign
of $\Delta t_{\sigma}=t_{\sigma} -t^0_{\sigma}$. 
Even though, for a given spin $\Delta t_{\sigma}$ can be negative,
the difference in the total kinetic
energies of the interacting  and the non-interacting system, 
$\Delta T= 
N [\Delta t_{\uparrow} +\Delta t_{\downarrow}]$, is positive
for any value of $\zeta$ and $r_s$.\cite{Holas87}
 
The negative value of $\Delta t_{\uparrow}$ for 
negative polarizations is easily understood if 
the variation with the polarization of the carrier
effective mass is considered. 
It is
well known that in a polarized electron gas the mass of the
minority (majority) spin populations increases (decreases) with
the value of the polarization.\cite{Overhauser71} 
Therefore, for large enough  values
of $\zeta$ the mass renormalization dominates, and the
kinetic energy of the  minority (majority) carriers
is reduced  (increased) with respect to its free electron value. 
Also, as $\zeta$ increases, the exchange effects between
the minority (majority) spins are greatly reduced (increased).
Due to the combination of these two effects 
the electrons from the majority spin population  will tend to accumulate
around the few minority spins increasing the local charge around
them. This generates a negative value 
of $G^{n}({\bf q}\rightarrow \infty)$ for the minority spins.
The opposite is true for the majority spins.

At finite values of the polarization, 
the kinetic energy of each spin population 
is dominated by the exchange contribution,
$\displaystyle \Delta t_{\sigma}=
-\frac{r_s}{2} \frac{\partial \epsilon^x }
{\partial r_s} - \epsilon^x_{\sigma} \sim  \zeta /r_s$, 
a behavior that is independent of the parameterization
used to obtain the spin-dependent correlation energy, 
Eq.~(\ref{eq:spincorrelation}) in our case.
At  large values of $r_s$,  the correlation effects
become more important as it can be seen on Fig.~\ref{fig:gammapol}.
Finally, we point out that the large effect of the
polarization in the value of $\gamma_{\uparrow}(r_s,\zeta)$ is mainly
due to the denominator $(1+\zeta)^2$ of Eq.~(\ref{eq:Gn}).

Fig.~\ref{fig:gammars} displays $ \gamma_{\uparrow}(r_s,\zeta)$
as a function of $r_s$ for several values of $\zeta$ 
between $\zeta=-0.5$ and $\zeta=1$.
The same trends as in Fig.~\ref{fig:gammapol} become apparent:
strong dependence on $\zeta$ and weak dependence with $r_s$.
For  values of $\zeta < 1$, 
$\gamma_{\uparrow}$  increases
with $r_s$, reaches a maximum and decreases afterwards. 
For example, for $\zeta=0$ the maximum is reached at $r_s=2.88$.
For $\zeta= 1$, $\gamma_{\uparrow}$ monotonically increases with $r_s$.

\section{\bf Spin Dependent Local Field Corrections}

Numerous parametrized expressions of the local field factors 
for the unpolarized three-\cite{sing70,vash72,zhu86,Corra98} 
and two-dimensional \cite{iwa91,bul96,Davoudi01}
electron gas has been suggested since the pioneering work of Hubbard. 
\cite{Hubb57} All these parametrized schemes rely on finding a
function of the electronic density that smoothly interpolates 
between the short and long wavelength limits of the local
field factors. 
However, parametrized expressions of $G^{\pm}$
for the polarized gas have not received so much attention.
Numerical calculations in the polarized electron gas are also 
computationally challenging and, so far, 
only results for the ground state energy
and the pair-distribution functions are available.
\cite{Tan89,Bulu02,Dav01,Moroni02}

As we have discussed in the previous section,
at small values of ${\bf q}$ the local factors follow a linear
dependence,
$\displaystyle G^{\pm}_{\sigma}({\bf q} \rightarrow 0)= 
\alpha^{\pm}_{\sigma}(r_s,\zeta)~ \tilde{q}$, where
$\tilde{q}={\bf q}/k_F$.
In the opposite end of the spectrum the local corrections 
follow a linear plus constant dependence,
$\displaystyle G^{\pm}_{\sigma}({\bf q} \rightarrow \infty)=
\beta^{\pm}(r_s)+ \gamma_{\sigma}(r_s, \zeta)~\tilde{q}$.
Given the diversity of behaviors displayed by the parameters 
$\alpha^{\pm}_{\sigma}$, $\beta^{\pm}$ and 
$\gamma_{\sigma}$,
we will consider a general
interpolating scheme for all values of $r_s$ and $\zeta$.
We parametrized $G^{\pm}$ with two fitting parameters, 
$q^{\pm}_0$ and $q^{\pm}_1$ as:
\be
G^{\pm}_{\sigma}({\bf q})= \alpha^{\pm}_{\sigma}   \tilde{q}
~e^{-(\tilde{q}/q^{\pm}_0)^4}
+ (\beta^{\pm}  + \gamma_{\sigma} \tilde{q})
[1-e^{-(\tilde{q}/q^{\pm}_1)^4}].
\label{eq:fitGn}
\ee

\begin{table}[t]
\begin{minipage}{\linewidth}
\caption{Optimal fit parameters for the local field factors as 
parametrized in Eqs.~(\ref{eq:fitGn}) and (\ref{eq:fitgaussian}).}
\begin{ruledtabular}
\begin{tabular} {||c|c|c|c|c|c||}
             & i=0 & i=1& & i=0 & i=1 \\
\hline $a^+_i$ &2.4847 &2.6523  & $a^-_i$ & 3.0578 & 3.9225 \\
\hline $b^+_i$ &0.6972 & 0.1927 & $b^-_i$ & 0.6844 & 0.6407  \\
\hline $c^+_i$ &0.4371 &0.1384  & $c^-_i$ & 1.2545 & 0.2507  \\
\end{tabular}
\label{tab:parameters}
\end{ruledtabular}
\end{minipage}
\end{table}

\begin{figure}
\begin{minipage}{\linewidth}
\includegraphics[width=\textwidth]{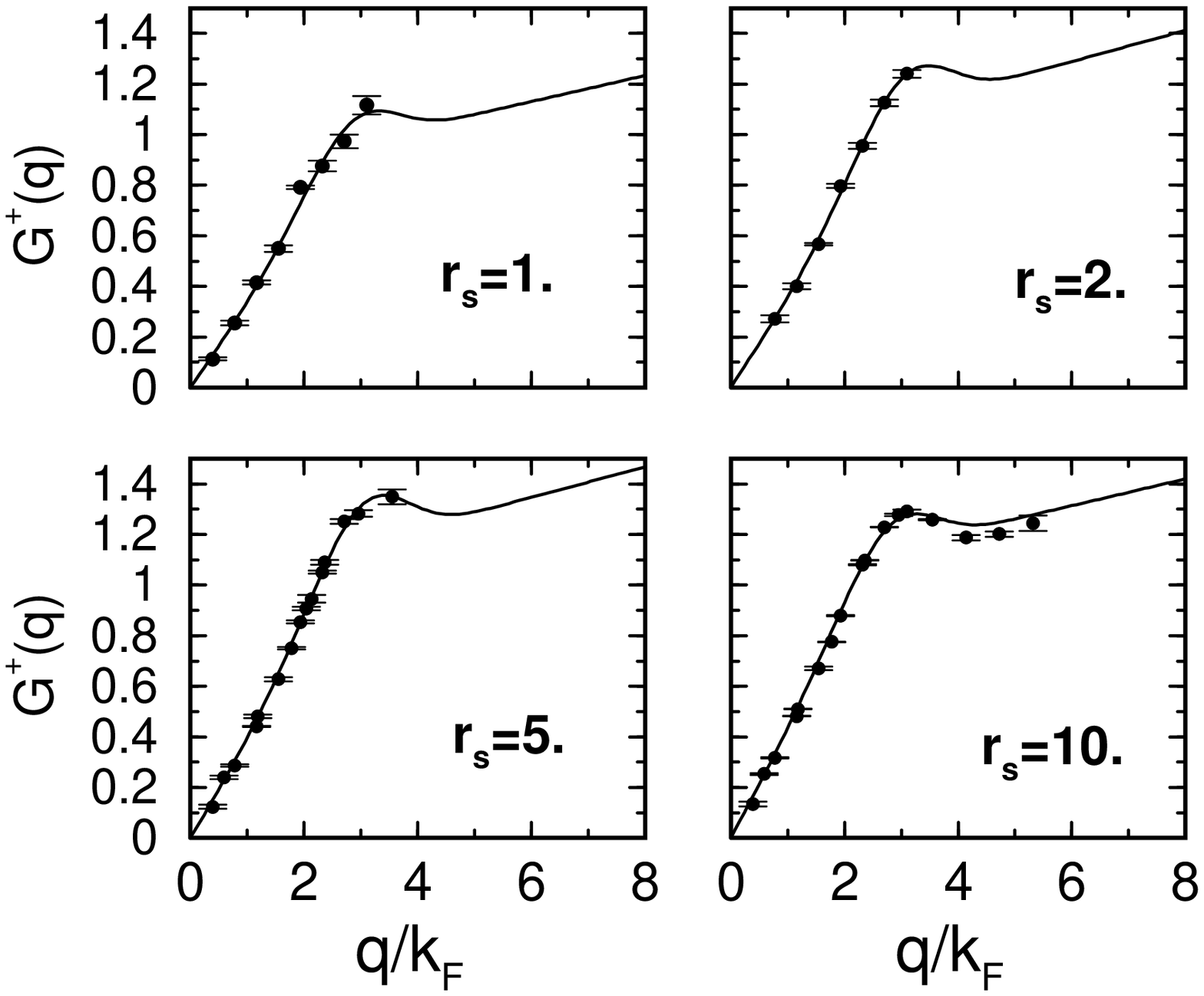}
\caption {Local field correction factor $G^+({\bf q},\omega=0)$ versus 
normalized momentum ${\bf q}/k_F$ for $r_s=1, 2, 5$ and $10$.
Black circles correspond to the diffusion Monte Carlo (DMC) results of 
Ref.~\onlinecite{DMC} and solid lines are calculated according
to Eqs.~(\ref{eq:fitGn}) and (\ref{eq:fitgaussian}).}
\label{fig:Moronigp}
\end{minipage}
\end{figure}

\begin{figure}
\begin{minipage}{\linewidth}
\includegraphics[width=\textwidth]{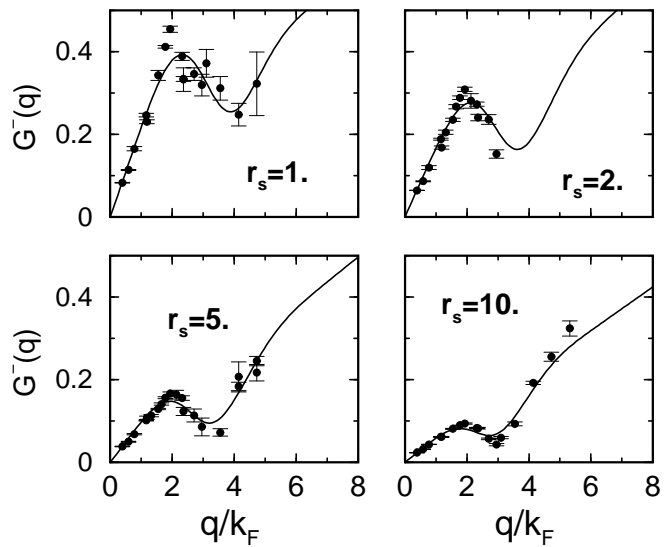}
\caption {Local field correction factor $G^-({\bf q},\omega=0)$ versus 
normalized momentum ${\bf q}/k_F$ for the same values of $r_s$.
Black circles correspond to the diffusion Monte Carlo (DMC) results of 
Ref.~\onlinecite{DMC} and solid lines are calculated according
to Eqs.~(\ref{eq:fitGn}) and (\ref{eq:fitgaussian}).}
\label{fig:Moronigm}
\end{minipage}
\end{figure}
  
The fast decreasing exponential factors 
$\exp{[-(\tilde{q}/q^{\pm}_0)^4]}$ and $\exp{[-(\tilde{q}/q^{\pm}_1)^4]}$
are needed to mimic the rapid evolution of the local functions from 
their small ${\bf q}$ to their large ${\bf q}$ limiting behaviors.
This fast evolution is clearly displayed on the  
latest diffusion Monte Carlo (DMC) results,\cite{DMC} 
which show how the local factors
$G^{+}({\bf q})$ and $G^{-}({\bf q})$
of the unpolarized gas
follow the predicted
linear dependence at small values of the wavevector and 
rapidly reach their asymptotic large wavevector limits at not very large
values of $\tilde q$.
Since there are only numerical results for the
unpolarized electron gas, we consider
the parameters $q^{\pm}_0$ and $q^{\pm}_1$ as functions only on 
the electron density. By fitting Eq.~(\ref{eq:fitGn}) to the 
DMC results of Ref.~\onlinecite{DMC} 
we find that the two fitting parameters are smoothly varying
functions of the coupling strength and can be parametrized as:
\be
q^{\pm}_i=\frac{a^{\pm}_i + b^{\pm}_i r_s + c^{\pm}_i r_s^{3/2}}
{1+\frac{1}{2}~c^{\pm}_i r_s^{3/2}},
\label{eq:fitgaussian}
\ee
where the parameters $a^{\pm}_i, b^{\pm}_i$ and $c^{\pm}_i$ 
are given in Table~\ref{tab:parameters}.

In Fig.~\ref{fig:Moronigp} we compared our results for 
$G^+({\bf q})$ of the unpolarized electron gas at $r_s=1,2,5$ and $10$, 
Eqs.~(\ref{eq:fitGn}) and (\ref{eq:fitgaussian}),
with the DMC results.\cite{DMC}
In Fig.~\ref{fig:Moronigm} our results for
$G^-({\bf q})$ of the unpolarized electron gas are also compared with
the numerical equivalent data. 
It is clear that our  parametrized expressions agree very well
with the Quantum Monte Carlo results.\cite{DMC}
By adding more free parameters 
to Eq.~(\ref{eq:fitGn}) the agreement with the numerical results will 
improve, but the question of how the parameters of the fit
evolve with the spin polarization remains, as yet, unanswered.
Thus, we keep the parameterization as simple as
possible: our fitting scheme, Eq.~(\ref{eq:fitGn}),
is the same
for both local functions, it contains only two free parameters and
these two  parameters have the same functional dependence with the
electronic density, Eq.~(\ref{eq:fitgaussian}).

\begin{figure}
\begin{minipage}{\linewidth}
\includegraphics[width=\textwidth]{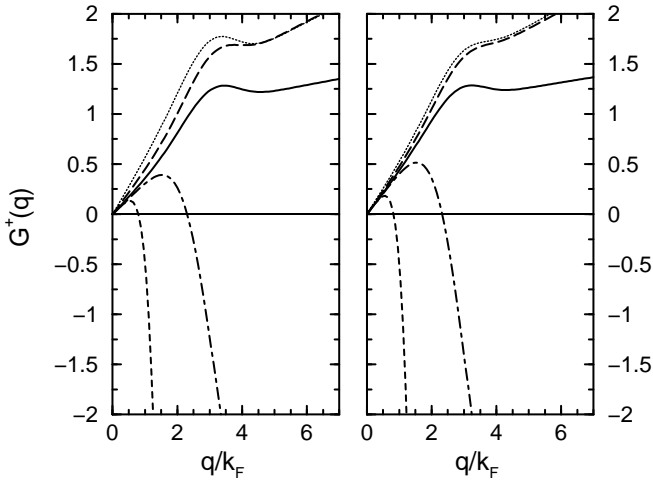}
\caption {Local field corrections $G^{+}_{\sigma}({\bf
q},\omega=0)$ versus normalized momentum ${\bf q}/k_F$ for a
two dimensional electron gas with $r_s=2$ (left panel) 
and $r_s=10$ (right panel). Results for $G^{+}$ at 
$\zeta=0$ (solid lines), $G^{+}_{\uparrow}$ (long-dashed lines) and 
$G^{+}_{\downarrow}$ (dot-dashed lines) at $\zeta=0.5$,
and $G^{+}_{\uparrow}$ (dotted lines) and 
$G^{+}_{\downarrow}$ (dashed lines) at $\zeta=0.9$ are displayed.}
\label{fig:Gplus}
\end{minipage}
\end{figure}

Fig.~\ref{fig:Gplus} shows the momentum dependence of the local
field functions $G^+_{\sigma}({\bf q})$ for 
three values of the polarization, $\zeta=0,0.5$ and $0.9$,
and two values of the coupling strength,  
$r_s=2$ (density of 7.43 $\cdot 10^{10}$ cm$^{-2}$ in GaAs)
and  $r_s=10$  (density of 0.297 $\cdot 10^{10}$ cm$^{-2}$ in GaAs).
The  factor associated with the majority spins, 
$G^+_{\uparrow}$, behaves quite different from the one of the 
minority spins, $G^+_{\downarrow}$. The field factor $G^+_{\uparrow}$
is always positive for positive values of the polarization. 
It slightly increases with the degree of polarization,
but it keeps the  characteristic peak of the 
unpolarized local factor around the same value of ${\bf q}$. 
This peak is a residue of the 
sharp peak in the exchange potential,\cite{Wang84} which is washed
out and/or shifted to higher values of ${\bf q}$
by the inclusion of short-range correlations.\cite{pede97}
On the other hand, the behavior of $G^+_{\downarrow}$ 
is dominated by the forced
change in its slope, from a positive value at small ${\bf q}$
to a large negative value at large wavevectors. 
As a result,
$G^+_{\downarrow}$ has always a maximum, whose position 
shifts to lower values of $\tilde q$ with increasing $\zeta$
and it is strongly dependent 
on the precise parameterization used. For example, 
it can shift to larger values of $\tilde q$
if $q^+_0$ and $q^+_1$ are also  functions of $\zeta$.
This issue should
be explored in greater detail. Finally, by comparing 
the field factors for $r_s=2$ and  $r_s=10$ we conclude 
that their dependence on $r_s$ is weak.

\begin{figure}
\begin{minipage}{\linewidth}
\includegraphics[width=\textwidth]{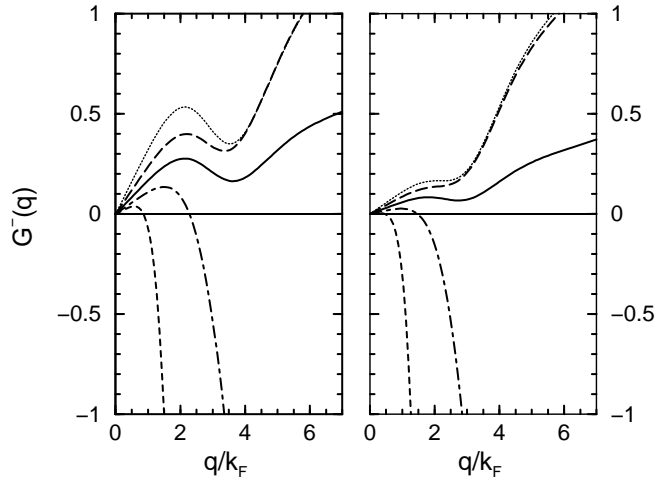}
\caption {Local field corrections $G^{-}_{\sigma}({\bf
q},\omega=0)$ versus normalized momentum ${\bf q}/k_F$ for a
two dimensional electron gas with $r_s=2$ (left panel) 
and $r_s=10$ (right panel). Results for $G^{-}$ at 
$\zeta=0$ (solid lines), $G^{-}_{\uparrow}$ (long-dashed lines) and 
$G^{-}_{\downarrow}$ (dot-dashed lines) at $\zeta=0.5$,
and $G^{-}_{\uparrow}$ (dotted lines) and 
$G^{-}_{\downarrow}$ (dashed lines) at $\zeta=0.9$ are included.}
\label{fig:Gminus}
\end{minipage}
\end{figure}

Fig.~\ref{fig:Gminus} displays the local field 
factor $G^-_{\sigma}({\bf q})$
versus normalized momentum for the same values of $\zeta$ and
$r_s$ used in Fig.~\ref{fig:Gplus}. 
The main difference
between  $G^+_{\uparrow}({\bf q})$ and $G^-_{\uparrow}({\bf q})$ 
is that the latest one displayed a sharper peak around
$\tilde{q} \sim 2$. It appears that higher order effects,
which are important in the computation of $G^+_{\uparrow}$,
cancel out in calculations of $G^-_{\uparrow}$ due to the
antisymmetric averaging over spin. \cite{Atwal02}
It is also noticeable how the local factors change
with increasing $r_s$. 

\section{\bf Conclusions}

We have considered an analytic parameterization of the spin dependent
local field factors of the polarized two-dimensional electron gas, 
Eqs.~(\ref{eq:fitGn}) and (\ref{eq:fitgaussian}). Our
parameterization incorporates the known asymptotic limits of the 
local corrections and gives an 
accurate fit of the available quantum Monte Carlo data.\cite{DMC}
We found that the local field
corrections associated to the minority spins strongly 
depend on the  polarization, while
the local field functions of the majority spins are less
affected by the degree of polarization. This is mainly due to
the negative value of the linear term  on the large ${\bf q}$ limit 
of the local field factors of the minority spins. 

The analytic parameterization used has only two  parameters 
which have been fitted to reproduce the latest DMC results
for the unpolarized electron gas.\cite{DMC}
Since, that there are not data on polarized systems 
we have considered these two  parameters  as functions only
on the electron density. Therefore, further study 
will be needed to evaluate their dependence on the spin polarization
and the efficacy of our parameterization at large values
of $\zeta$.

In conclusion,
we believe that our approach  provides a realistic qualitative 
description of the 
paramagnetic phase of the polarized electron gas.
Caution, however, should be
exercised in applying our calculation in the limit of $\zeta$
approaching unity, where the paramagnetic model breaks down.
We have found that for small values of $r_s$ and 
large values of  $\zeta$, the magnetic susceptibility and
the inverse dielectric constant develop a pole at the
same value of the electronic density and the spin polarization.
This fact signals a charge-spin density wave instability
in the polarized electron gas and it will be discussed elsewhere. 
\cite{Moreno03}

 {\bf Acknowledgments}

We are grateful to Dr. Moroni and Dr. Senatore for providing
us with the results of their numerical calculation of 
the local field factors in the unpolarized electron gas.
We gratefully acknowledge the financial support provided by the
Department of Energy, grant no. DE-FG02-01ER45897.


\begin{thebibliography}{10}
\bibitem{unpol-pol} In an unpolarized electron gas, 
$G^{+}$ appears only in the expressions of the
electrical response functions, and
$G^{-}$ only on the magnetic response susceptibilities.
However, in a polarized electron system the four local field factors 
appear in all the response functions.
\bibitem{numbers} Where $a^{*}_B$ is 
the effective Bohr radius of the
system, $a^{*}_B=\hbar^2/(m^* e^{*2})$.
For a regular three dimensional metal
$a^{*}_B=a_0=0.529~\cdot 10^{-8}$ cm 
and $a^{*}_B= 195 \hspace{0.1cm} a_0= 1.04~\cdot10^{-6}$ cm for GaAs.
\bibitem{approximation} The quenching of the divergence 
at $\zeta \pm 1$ will depend on the precise parameterization used 
for the correlation energy.
\end{thebibliography}
\end{document}